%% file: template.tex
\documentclass[journal]{vgtc}                % final (journal style)
\ifpdf%                                % if we use pdflatex
  \pdfoutput=1\relax                   % create PDFs from pdfLaTeX
  \usepackage{graphicx}                % allow us to embed graphics files
  \DeclareGraphicsExtensions{.pdf,.png,.jpg,.jpeg} % for pdflatex we expect .pdf, .png, or .jpg files
\else%                                 % else we use pure latex
  \usepackage{graphicx}                % allow us to embed graphics files
  \DeclareGraphicsExtensions{.eps}     % for pure latex we expect eps files
\fi%

\graphicspath{{figures/}{pictures/}{images/}{./}} % where to search for the images

\usepackage{microtype}                 % use micro-typography (slightly more compact, better to read)
\PassOptionsToPackage{warn}{textcomp}  % to address font issues with \textrightarrow
\usepackage{textcomp}                  % use better special symbols
\usepackage{mathptmx}                  % use matching math font
\usepackage{times}                     % we use Times as the main font
\usepackage{cite}                      % needed to automatically sort the references
\usepackage{tabu}                      % only used for the table example
\usepackage{booktabs}                  % only used for the table example
\usepackage{xcolor}
\usepackage{bm}
\usepackage{amsmath}
\usepackage{amsfonts}
\usepackage{amsthm}
\usepackage{algpseudocode} 
\usepackage{algorithm}
\usepackage[hang,flushmargin]{footmisc}
\usepackage{enumitem}
\onlineid{1081}

\vgtccategory{System}
\vgtcpapertype{application/design study}

\title{Auditing the Sensitivity of Graph-based Ranking with Visual Analytics}

\author{Tiankai Xie, Yuxin Ma, Hanghang Tong, My T. Thai, Ross Maciejewski}
\authorfooter{
\item
\vspace{-1.2mm}
T. Xie, Y. Ma,  and R. Maciejewski are with Arizona State University. E-mail: \{txie21, yuxinma, rmacieje\}@asu.edu.
\item
H. Tong is with the University of Illinois at Urbana-Champaign. E-mail: htong@illinois.edu.
\item
M. Thai is with the University of Florida. E-mail: mythai@cise.ufl.edu.
}

\abstract{Graph mining plays a pivotal role across a number of disciplines, and a variety of algorithms have been developed to answer who/what type questions. For example, what items shall we recommend to a given user on an e-commerce platform? The answers to such questions are typically returned in the form of a ranked list, and graph-based ranking methods are widely used in industrial information retrieval settings. However, these ranking algorithms have a variety of sensitivities, and even small changes in rank can lead to vast reductions in product sales and page hits. As such, there is a need for tools and methods that can help model developers and analysts explore the sensitivities of graph ranking algorithms with respect to perturbations within the graph structure. In this paper, we present a visual analytics framework for explaining and exploring the sensitivity of \textbf{any} graph-based ranking algorithm by performing perturbation-based what-if analysis. We demonstrate our framework through three case studies inspecting the sensitivity of two classic graph-based ranking algorithms (PageRank and HITS) as applied to rankings in political news media and social networks.
} % end of abstract


\teaser{
  \centering
  \includegraphics[width=\linewidth]{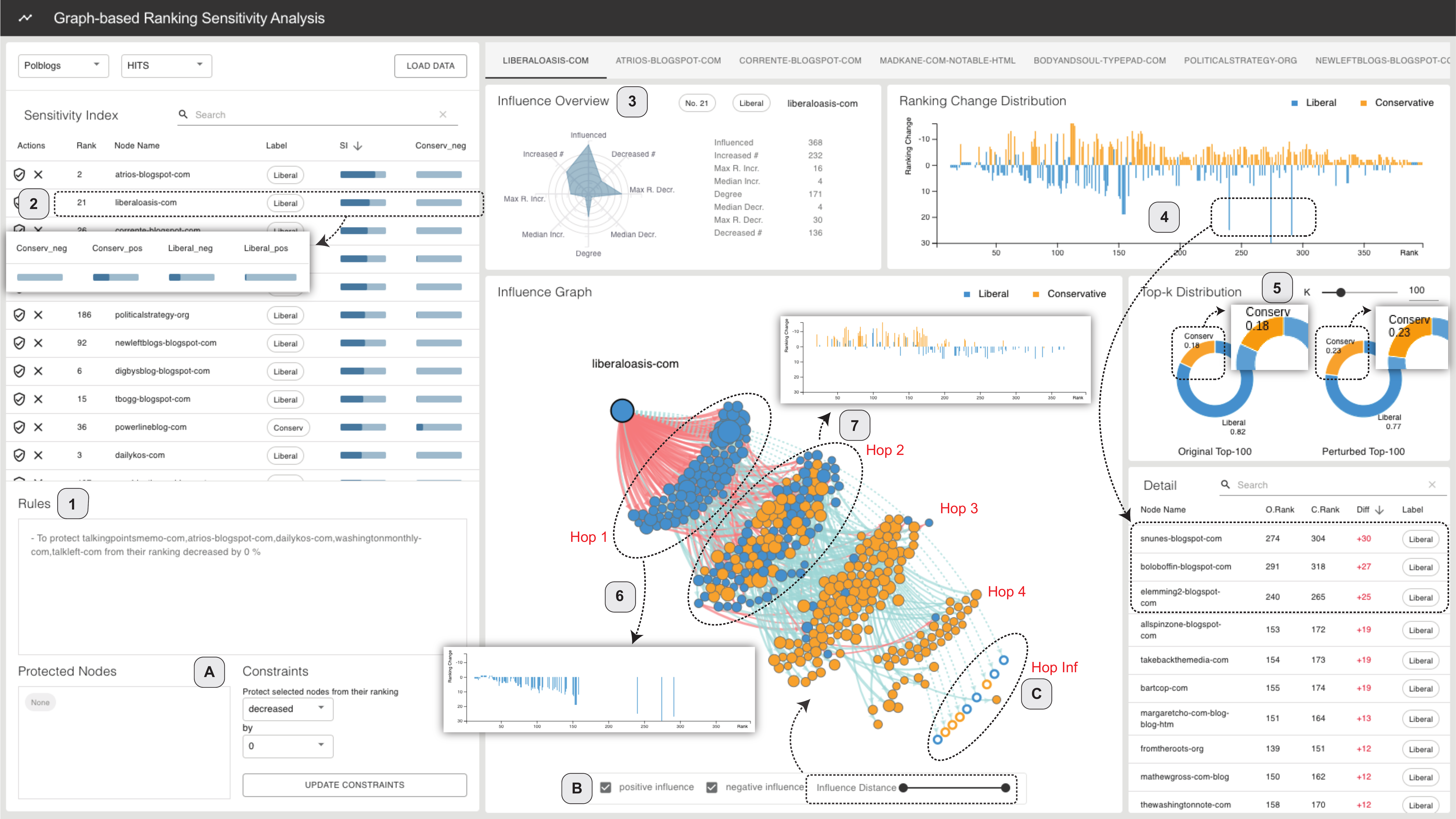}
  \caption{Sensitivity analysis of HITS on political blogs. (1) A predefined rule is set to exclude all perturbations that would cause the rankings of the top-5 blogs to decrease. (2) The blog \texttt{liberaloasis.com} has the largest influence under this constraint, and its removal can increase the rankings of the conservative blogs while decreasing the rankings of the liberal blogs. (3) The influence overview indicates that nearly 2/3 of the influenced nodes see a ranking increase. (4) The ranking change distribution view further shows that most of the ranking-increased nodes are conservative blogs and most of the ranking-decreased nodes are liberal blogs, from which the top-3 heavily influenced nodes are ranked 200th or below. (5) The top-$k$ proportional view shows that the proportion of liberal blogs decreased from 82\% to 77\% in the top-100 due to the perturbation.(6, 7) The influence graph view implies that the removal of \texttt{liberaloasis.com} has a direct influence on the majority of the liberal nodes (including the top-3 influenced nodes), and as the influence distance increases, more conservative nodes are indirectly influenced.}
	\label{fig:teaser}
}

\vgtcinsertpkg

\begin{document}



\input{1_introduction}

\input{2_related_work}
\input{3_design_overview}
\input{4_framework}

\input{5_case_study}

\input{6_conclusion}

\acknowledgments{
This work was supported by the U.S. Department of Homeland Security under Grant Award 2017-ST-061-QA0001 and 17STQAC00001-03-03, and the National Science Foundation Program on Fairness in AI in collaboration with Amazon under award No. 1939725. The views and conclusions contained in this document are those of the authors and should not be interpreted as representing the official policies, either expressed or implied, of the U.S. Department of Homeland Security.
}

\bibliographystyle{abbrv-doi}

\bibliography{template}
\end{document}

%% file: 1_introduction.tex
\maketitle

\section{Introduction}

Currently, the development of visual analytics methods and tools for explainable artificial intelligence (XAI) primarily tackles analytical tasks in vector-space learning, such as classification~\cite{Alsallakh2014,Zhang_2018_CVPR}, clustering~\cite{8017580,8440035}, and outlier detection~\cite{8017580,10.1145/3173574.3174237}. However, graph-based learning algorithms are significantly different from vector-space representations, and these differences have not been sufficiently studied in the visual analytics community. Specifically, graph-based ranking algorithms have received little attention; however, algorithms such as PageRank~\cite{ilprints422} and HITS~\cite{Kleinberg99authoritativesources} are foundational in industrial information retrieval settings. In these information retrieval settings, a person searches a graph-based dataset looking for relevant objects, and the resultant ranking order has a major impact with respect to exposure. For example, recent work by Singh and Joachims~\cite{10.1145/3219819.3220088} demonstrated that the exposure of resumes to potential employers could be reduced by upwards of 30\% if an item's rank fell by as little as three places.   

Previous work~\cite{Kang:2019:NND:3357384.3357910, ng2001link, 10.1137/090772745, kang2018aurora} has demonstrated that the results of such graph-ranking algorithms can be highly sensitive to perturbations within the graph structure, and these sensitivity issues give rise to ranking manipulations.
Given the importance of the ranking results, it is imperative that algorithm designers and analysts understand the underlying algorithmic sensitivities and vulnerabilities.
Consider a news navigation website~\cite{adamic2005political} where the consumer can search political-related blogs and posts. The search result rankings are determined by a graph ranking algorithm, and higher ranked stories are more likely to be read and shared. Here, one could imagine a nation-state actor that would want to promote biased content. The nation-state actor can create webpages to add various links in the graph structure, or even identify websites to shadow ban, which could manipulate the ranking results so that certain political opinions are more exposed to the public. 
Given the importance of such rankings, it is critical that model developers have access to tools that can support them in understanding the ranking methods' sensitivity to structural changes in the graph. 

In this paper, we propose a modularized visual analytics framework that facilitates auditing and diagnosing \textbf{any} graph-based ranking method's sensitivities by performing what-if analysis over a given graph dataset via node perturbation. 
The interactive perturbation of a graph's nodes through coordinated views enables analysts to explore and identify algorithmic sensitivities. A summarization view for the sensitivity index (i.e., the degree of the ranking method's sensitivity to the perturbation) facilitates the identification of the graph-ranking method's instance-level sensitivity. A group of views quantifies the impacts of perturbations through the comparison of statistical information about the ranking results, and a local graph influence view supports the inspection of ranking changes due to changes in the graph topology.
To demonstrate our framework, we explore three case studies using real-world datasets including: a Facebook social circle~\cite{leskovec2012learning}, Political blogs~\cite{adamic2005political}, 
and a Reddit interaction network~ \cite{10.1145/3178876.3186141}.
While our framework is designed to support any general graph-based ranking algorithm, only PageRank and HITS are used for demonstration purposes. 
Our contributions include:
\begin{itemize}[topsep=0.5em,partopsep=0em,parsep=0.2em,itemsep=0em,leftmargin=*,labelsep=0.4em]
  \item A visual analytics framework that facilitates performing what-if analysis on graph data to reveal the instance-level sensitivity in terms of \textbf{any} graph-based ranking method, and;
  \item A novel representation of the influence graph caused by specific perturbations to illustrate the relationship between ranking influence and topological structures.
\end{itemize}

%% file: 2_related_work.tex
\section{Related Work}
Our work focuses on explaining the sensitivities of graph-based ranking models in relationship to perturbations in the graph structure. In this section, we review recent work on ranking tasks and sensitivities in graph theory and visualization methods applied to graphs and rankings.

\subsection{Graph Ranking}

Graph ranking methods are ubiquitous, with applications of these algorithms found in web searches, e-commerce, hiring, and numerous other domains.
The most well-known methods for graph ranking include Google's PageRank~\cite{ilprints422} and Kleinberg's HITS~\cite{Kleinberg99authoritativesources}. 
PageRank computes the quality of hyperlinks to webpages in order to approximate the importance of webpages. Webpages are treated as nodes of a graph, and the edges of this graph are the hyperlinks between webpages. Given a graph \textbf{G} with $n$ nodes, PageRank computes the importance of nodes as:
\begin{equation}
\label{eq: pagerank}
\begin{split}
    \bm{r} = c\bm{Ar} + (1 - c)\bm{t}
\end{split}
\end{equation}
where \textbf{r} is a vector of size $n$ that denotes the \textit{PageRank Value (or PR Value)} for each node in the graph \textbf{G}. The higher the PR value of a node, the higher ranked (or more important) the node is. \textbf{A} is the normalized adjacency matrix of graph \textbf{G}, $c$ is a constant damping factor, which is typically set as 0.85~\cite{brin1998anatomy}, and \textbf{t} is the teleportation vector, representing the initial PR value for each node. \textbf{t} is typically the uniform probability distribution $\frac{1}{n}$ \textbf{1}. The computation of \textbf{r} will eventually converge at the final ranking value for each node.

Unlike PageRank, HITS (Hyperlinked Induced Topic Search)~\cite{Kleinberg99authoritativesources}
treats webpages as ``authorities'' and ``hubs'', where ``authorities'' denote the webpages with more incoming hyperlinks, and ``hubs'' denote the webpages with more outgoing hyperlinks. Two update rules are applied to compute each node's ``authority'' score and ``hub'' score:
\begin{equation}
\begin{cases}
  auth(v) = \sum_{w \in V_{to}} hub(w) \\    
  hub(v) = \sum_{w \in V_{from}} auth(w)
\end{cases}
\end{equation}
where $V_{to}$ is a set of all nodes which point to the node $v$, and $V_{from}$ is a set of all nodes that node $v$ points to. The authority and the hub vector are initialized to 1 for each node, and the two computations are performed repeatedly until the values of the two vectors converge.

Both PageRank and HITS are propagation-walked-based techniques, and there are numerous variants of these algorithms. For example, the Personalized PageRank algorithm~\cite{ilprints422} initializes a biased  teleportation vector instead of one that has the same teleportation value for all nodes in a graph. The biased teleportation vector enables the generation of distinct ranking results that can be personalized for a set of nodes that the vector corresponds to.  ItemRank~\cite{Gori:2007:IRB:1625275.1625720} replaces the adjacency matrix with the stochastic matrix of a graph to transform PageRank into a biased version, which is commonly applied in recommendation systems. IsoRank~ \cite{singh2007pairwise} treats the task of comparing sets of graphs to find correspondences between nodes as the eigenvalue problem and has been used to align protein to protein interaction networks. TwitterRank \cite{Weng:2010:TFT:1718487.1718520} utilizes a transition probability matrix in which the similarity between twitterers on certain topics is used to identify the topic-sensitive influential twitterers, and TopicRank~ \cite{bougouin-etal-2013-topicrank} uses the semantic relationship between topics as part of the document ranking process. These methods all have the same underlying data structure requirements, which enables our framework to seamlessly swap between algorithms.

\subsection{Graph Auditing}
While numerous graph-based ranking algorithms have been developed, it is only recently that researchers have begun exploring methods to audit network/graph learning methods in an attempt to identify issues of fairness and bias.
Kang et al.~\cite{kang2018aurora} define the \textit{PageRank Auditing Problem} as the task of finding the $k$ graph elements (nodes or edges)  that have the largest influence on the overall ranking changes of a given graph. Kang et al. measure the ranking changes of a given graph to compute the derivative of a loss function of the Pagerank vector over the adjacency matrix. However, in this method, the $k$ graph elements have to be computed $k$ times in total, and each time the ranking method has to be rerun and the derivative values must be re-computed. 

Other sensitivity analysis methods have focused on performing perturbations on graph elements to explore vulnerabilities of graph-based learning models~\cite{Sun2018, chen2018fast, 10.1145/3219819.3220078, dai2018adversarial, chen2018link}.
For example, Ng et al.~\cite{ng2001link} utilize the techniques in matrix perturbation theory and coupled Markov chain theory to evaluate the stability of PageRank and HITS. Chartier et al.~\cite{10.1137/090772745} perform a comparative analysis of the effects of perturbations on graph structures exploring various ranking methods. 
However, such definitions of auditing only determine \textbf{which} elements have a high influence for a specific ranking method. They do not provide visual details on \textbf{why} and \textbf{how} the perturbations influence ranking. Our framework is designed to support the analysis of multiple perturbations to help explain potential vulnerabilities in graph-based ranking algorithms. 

\subsection{Graph Drawing and Ranking Visualization}
A key mechanism for supporting the analysis of graph-based ranking algorithms is visualization.
Graph drawing methods focus on algorithms that optimize the layout of the graph structure to highlight key features in a graph. Numerous graph drawing methods have been developed (see recent state-of-the-art reports for a comprehensive summary~\cite{beck2017taxonomy, beck2014state, von2011visual}), with recent methods focusing on exploiting deep learning approaches to generate graph layouts for large datasets~~\cite{Kwon_2019, Wang2019}, edge bundling to improve layout readability~\cite{Wang_Bundle2019, Sun2019}, and extensions to the force-directed layout~\cite{Zheng2019}.
In this framework, we apply a customized radial graph layout with redesigned encodings to reveal the effects of perturbation.

Another major task in our proposed framework is to reveal the relationships between a graph node's ranking change and its topological structure, and many approaches have been proposed to visually represent ranking changes. For example, a variety of systems~\cite{Lu2015,Shi2012, vuillemot2015investigating, xia2017visualizing} have been developed to support visualizing ranking changes in time-series data. Lu et al.~\cite{Lu2015} explore the use of parallel coordinate plots to visualize ranking changes through time. Shi et al.~\cite{Shi2012} use a stacked area chart where the order shows the ranking and the stacks show the proportional changes. Vuillemot et al. combine a table view and a line chart to show ranking changes over time for a group of tabular data. Similarly, Xia et al.~\cite{xia2017visualizing} show the PageRank results of top Wikipedia topics and reveals the connection between those topics. 

Other research~\cite{Kwon2018,Weng2019,Gratzl2013} has focused on ranking and multi-dimensional data comparisons. Clustervision~\cite{Kwon2018} provides a proportional bar and radar glyph to show metrics for each cluster in a ranking list. SRVis~\cite{Weng2019} integrates spatial data into ranking visualization to support decision making. LineUp~\cite{Gratzl2013} utilizes bar charts to de-factorize and combine attributes to show the rank changes under selected attributes. Podium~\cite{Wall2018} uses an SVM classifier to help analysts rank the tabular data to fit their mental model, and GUIRO~\cite{8807245} is used to improve matrix reordering methods with animated reordering representations. However, none of those approaches explore the relationship between ranking changes and the topological structure of graphs.

\begin{figure*}[tbh]
    \centering	
    \includegraphics[width=2.00\columnwidth]{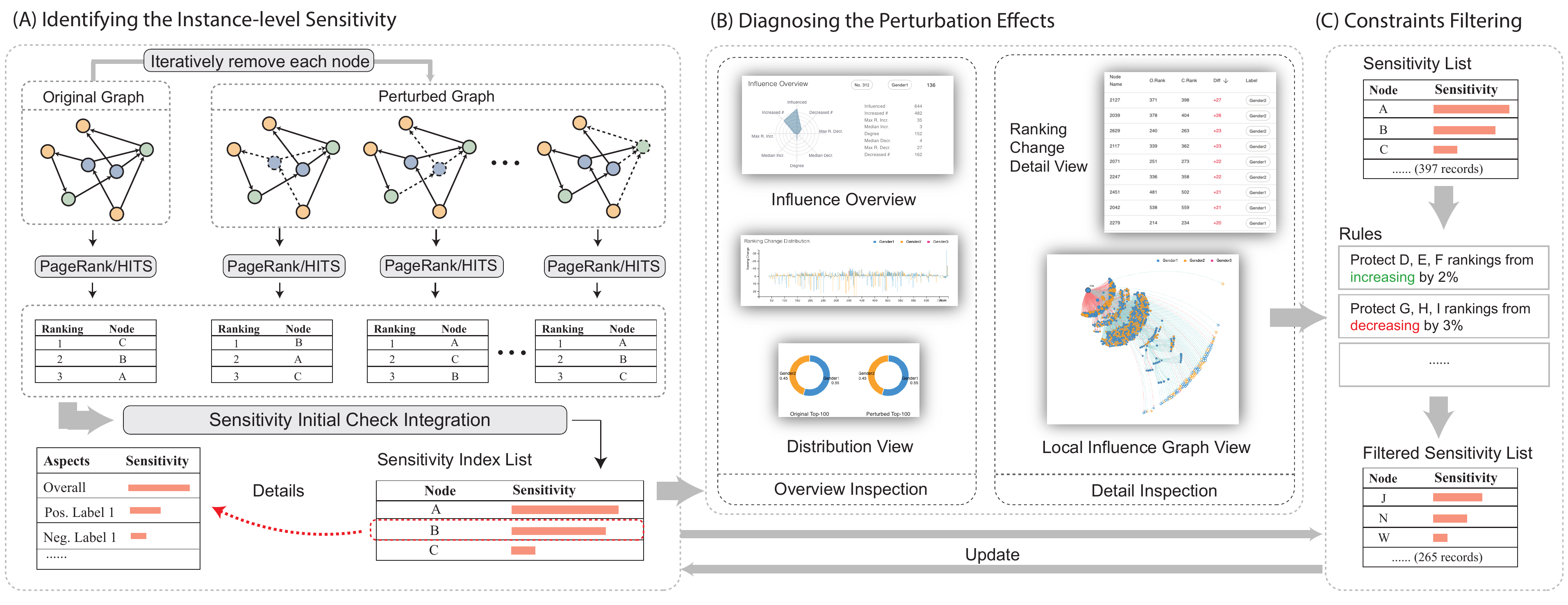}
    \caption{A visual analytics framework for identifying, auditing, and diagnosing a ranking method's sensitivity to instance-level perturbations. The framework consists of Identifying the Instance-level Sensitivity, Diagnosing the Perturbation Effects, and Constraints Filtering. (A) Perturbation is applied for each node instance. The ranking methods are rerun for each new graph generated. Sensitivity to the ranking method for each node's removal is calculated. (B) The analyst can explore one of the nodes to see the perturbation effects through both overview inspection and detailed inspection. (C) Then the analyst can apply constraints to the sensitivity index list based on their findings, and the sensitivity index list is updated. }
    \label{fig:visual_analytics_framework}
	\vspace{-4mm}
\end{figure*}

%% file: 3_design_overview.tex
\section{Design Overview}
Given the large scale use of graph-based ranking algorithms, we have developed a visual analytics framework to support developers and analysts in exploring and explaining ranking sensitivities. Our framework is designed to be robust to general graph-based ranking algorithms, and supports the removal of nodes as the key perturbation method. While other types of perturbations exist, such as adding/removing nodes/edges\cite{jia2020certified, wang2018attack, kang2018aurora}, our framework focuses on the removal of a node and its corresponding edges as a proof-of-concept interaction. The removal perturbation allows us to constrain the computational and exploration space. However, the interactions and design are robust to all types of perturbations, which will be explored in future work.
Our potential target audience includes researchers, developers, and analysts who are building and/or deploying graph-based ranking applications. Our goal is to facilitate those experts’ analysis of the sensitivity of their chosen ranking methods and support them in auditing the applied ranking algorithms before deployment.

\subsection{Analytical Tasks}
After reviewing recent literature on graph auditing~\cite{kang2018aurora, Kang:2019:NND:3357384.3357910}, we extracted common high-level tasks for the auditing process. These tasks were refined with our co-authors, domain-experts in graph-mining.

\vspace{2mm} \noindent \textbf{T1 Summarize instance-level sensitivity.} 
The key analytic task is to identify an individual node's ranking and the sensitivity of this node's ranking to changes in the graph structure. 
Our framework is designed to provide an overview of the ranking results, and enables analysts to explore any node's sensitivity to perturbation. For example:
\begin{itemize}[topsep=0.5em,partopsep=0em,parsep=0.2em,itemsep=0em,leftmargin=*,labelsep=0.4em]

    \item 
    \textbf{T1.1} Which perturbation causes the largest ranking changes?

    \item
    \textbf{T1.2} How do perturbations cause ranking changes, i.e., are there topological features that leading to ranking instability? 

    \item
    \textbf{T1.3} Are nodes with specific attributes more sensitive to perturbations in the network, i.e., does the removal of a node of group A lead to changes in the ranking of nodes in group B, where groups are defined by some underlying network attribute.
\end{itemize}

\vspace{2mm} \noindent \textbf{T2 Diagnose the perturbation effects.} 
What-if analysis~\cite{wexler2019the} has previously been used for XAI as a mechanism to investigate machine learning model performance for a range of data features. 
In our framework, we adopt this idea of what-if analysis to measure the output of any graph-based ranking algorithm by perturbing the input. 
This enables model developers to measure and explore the ranking changes and corresponding effects.
As we focus on \textbf{removing nodes} as the perturbation mechanism, a key analytical task is to support diagnosing changes caused by node removals including:

\begin{itemize}[topsep=0.5em,partopsep=0em,parsep=0.2em,itemsep=0em,leftmargin=*,labelsep=0.4em]

    \item \textbf{T2.1 Summarize the ranking influence of perturbation} The system should provide a summary of how perturbation has impacted the graph-based ranking results. 
    
    \item \textbf{T2.2 Enable the ranking influence comparison between subgroups}
    Each graph node represents an instance and may have attributes/labels that can be used to define class membership. 
    Questions about ranking changes are often strongly tied to questions of fairness related to graph attributes, for example, given a hiring database, are the ranking of female applicants more sensitive to changes in the graph structure than male applicants?
        
    \item \textbf{T2.3 Identify the topological influence caused by perturbations} The system also needs to support analysts in exploring how perturbations have influenced the graph topology.  

\end{itemize}

\vspace{2mm} \noindent \textbf{T3 Enable progressive analysis.} The system should support the analysis of multiple perturbations as analysts explore what-if scenarios. 

\subsection{Design Requirements}
 From the task requirements, we engaged in an agile design process with our domain experts, iterating over various visualization and interaction designs. Based on our discussions, prototyping and feedback, we have mapped different analytic tasks to a set of design requirements. 

\vspace{2mm} \noindent \textbf{D1 Visualize the Instance-level Sensitivity.}
The system should visualize ranking and auditing results for all instances (T1). The view for summarizing the instance-level sensitivity should include the sensitivity index (T1.1) for all nodes with respect to the node attributes (T1.3).

\vspace{2mm} \noindent \textbf{D2 Visualize the Effect of Perturbation.}
The system should be able to guide analysts to explore the perturbation effect of certain node's removal and support interactions such as sorting, searching and filtering to inspect the auditing results and corresponding perturbation effects (T2, T3). This view should include:
    
\begin{itemize}[topsep=0.5em,partopsep=0em,parsep=0.2em,itemsep=0em,leftmargin=*,labelsep=0.4em]
\item \textit{D2.1 Influence Overview}, which summarizes the perturbation's influence, the degree of ranking changes, and the proportion of nodes whose rankings are increased/decreased, etc. (T1.2, T2.1, T3)

\item \textit{D2.2 Distribution View}, which shows how the ranking position changes caused by the perturbation are distributed for each instance and the ranking distribution for each group of nodes. (T2.2, T1.3)

\item \textit{D2.3 Ranking Change Detail View}, which lists the influenced nodes for this perturbation. The view should support basic query operations, e.g., sorting, filtering and searching, etc.(T2.1, T3)

\item \textit{D2.4 Local Influence Graph View}, which illustrates the relationship between the ranking changes of nodes and the topological changes caused by the perturbation. (T2.3, T3)
\end{itemize}

%% file: 4_framework.tex
\section{Visual Analytics Framework}
Based on the analytic tasks and design requirements, we have developed a visual analytics framework (Figure~\ref{fig:visual_analytics_framework}) for auditing, diagnosing, and analyzing graph-based ranking methods' sensitivities to instance-level perturbation through what-if analysis. 
The framework is designed to first compute a sensitivity calculation for each node of the graph and integrate the results to form a list of all sensitivity information (Figure~\ref{fig:visual_analytics_framework} (A)). Once the precomputation is loaded, the analyst interacts with the system by choosing nodes to perturb, and the framework calculates and visualizes the perturbation effects (Figure~\ref{fig:visual_analytics_framework} (B)). The framework also supports filtering the list based on the analyst-defined rules to support the inspection of the data under a variety of constraints (Figure~\ref{fig:visual_analytics_framework} (C)). By supporting an iterative process of perturbing nodes and adding analyst-defined rules, this framework enables the auditing of the sensitivity of graph-based ranking algorithms.

The framework supports three main activities: instance-level sensitivity identification, perturbation diagnosis, and customized constraints filtering. Through instance-level sensitivity identification, the analyst can explore an overview of ranking sensitivity with respect to perturbation (node removal).
In the perturbation diagnosis, effects caused by the perturbation are displayed with respect to the statistical distribution, top-$k$ distribution, and influenced paths. From the detailed influence view, the analyst can identify the potential constraints and further apply those constraints to filter the results.
The analyst can repeat this process until they identify nodes of interest. Our framework is designed to be modular, enabling model designers to integrate any graph-based ranking algorithm. 
However, for discussion and demonstration purposes, we only explore PageRank and HITS.

\subsection{Identifying the Instance-level Sensitivity}
The first component of our framework is designed to support instance-level sensitivity analysis (or auditing) for nodes in the graph.
Kang et al.~\cite{kang2018aurora} defined sensitivity auditing as finding the $k$ graph elements (nodes or edges) that have the largest influence on the overall ranking changes of a given graph. 
This approach identifies the most influential element, removes this element, updates the graph structure, identifies the most influential element from the new graph structure, and continues repeating this process until $k$ elements are found.
While such a process is useful for identifying the most sensitive nodes, it does not directly incorporate a mechanism for measuring sensitivity for an individual node. 
In order to define sensitivity per node, we modify the definition of sensitivity auditing from Kang et al.~\cite{kang2018aurora}.
\newtheorem{definition}{Definition}
\begin{definition}
Given a graph $G$, an element (a node or an edge) to be removed $el_{rm}$ and a graph ranking method $f$, the \textbf{graph ranking sensitivity auditing} can be defined as finding the \textbf{sensitivity index} for each graph element of $G$. We denote the sensitivity index of element $el_{rm}$ as the degree of ranking method $f$'s sensitivity to the perturbation caused by the removal of this element $el_{rm}$. The sensitivity index of this element $el_{rm}$ is represented as:
\begin{equation}
\begin{split}
    s[el_{rm}] = sen(f, el_{rm}) = L(rp, rp') 
\label{eq: sensitivity_index}
\end{split}
\end{equation}
where $rp$ and $rp'$ denote the ranking positions for each node before and after the perturbation  respectively ($el_{rm}$ is not included in both $rp$ and $rp'$), and $L$ stands for a generic difference/distance measure between the ranking vectors before and after perturbation. \textbf{s} as the result represents the vector that contains the sensitivity index for every instance, and $s[el_{rm}]$ denotes the sensitivity index of $el_{rm}$.
\label{def: graph-ranking-sensitivity-auditing}
\end{definition}
Our proposed sensitivity index is used to help analysts compare sensitivities across removals of graph elements. As we focus on node removal in this work, $el_{rm}$ can be replaced by $v_{rm}$, which denotes the removed node. While there are many metrics available for calculating $L$, we apply the $L_1$ norm as the sensitivity metric as it directly measures the accumulated ranking position changes over all nodes.
Figure~\ref{fig:visual_analytics_framework} (A) illustrates how the Instance-level sensitivity module performs the initial sensitivity index check on each instance. 
Applying Definition 1, our framework first calculates the ranking change for every node by removing each node (and its corresponding edges) from the graph and calculates the ranking method on the perturbed graph. 
The removal of a node may cause the ranking of other nodes to change in both ranking directions (increase or decrease).
As such, the sensitivity can be summarized in multiple ways. 
For example, we could compute the overall positive/negative ranking change (or influence) that occurs when removing a node could be summarized or the class-specific positive/negative influence, where classes of nodes are defined based on their attributes and labels. 
In our framework, we calculate both a positive sensitivity index $sen_p$ and a negative sensitivity index $sen_n$ with respect to class labels as follows:
\begin{equation}
s_{pos}^b[v_{rm}] = sen_p(f, v_{rm}, b) = \begin{cases}
    \sum |rp_v - rp'_v|, rp_v - rp'_v > 0 \\
    0, otherwise 
\label{eq: positive_influence}
\end{cases}
\end{equation}
\begin{equation}
s_{neg}^b[v_{rm}] = sen_n(f, v_{rm}, b) = \begin{cases}
    \sum |rp_v - rp'_v|, rp_v - rp'_v < 0 \\
    0, otherwise 
\label{eq: negative_influence}
\end{cases}
\end{equation}
where $v \in V \wedge v \neq v_{rm} \wedge label(v) = b$, $V$ is the set of all nodes in the graph, $b$ is the class label, and $rp_v$ and $rp'_v$ are the ranking positions of node $v$ before and after the perturbation respectively. $\mathbf{s_{pos}}$ and $\mathbf{s_{neg}}$ represent the vectors of positive and negative sensitive index for each instance, in which $s_{pos}[v_{rm}] = \sum_{b \in B}s_{pos}^b[v_{rm}]$ and $s_{neg}[v_{rm}] = \sum_{b \in B}s_{neg}^b[v_{rm}]$, where $B$ is a set of labels. 
Eq.~\ref{eq: sensitivity_index}, Eq.~\ref{eq: positive_influence} and Eq.~\ref{eq: negative_influence}, are applied in Algorithm~\ref{alg:sensitivity_index_initial_check} to realize the the initial sensitivity index check for every node and every available label. The system then visualizes the output in the sortable sensitivity index list (D1), which
shows each node's current ranking and sensitivity indices with respect to the node's class label(s). 

\subsection{Diagnosing the Perturbation Effects}
The second component of our framework is designed to support what-if analysis. 
In this module, analysts begin by selecting a node from the sensitivity index list to further explore the perturbation effects. 
Several linked views are deployed: 

\setlength\textfloatsep{1.1mm}
\begin{algorithm}
	\caption{Sensitivity Index Initial Check}
	\label{alg:sensitivity_index_initial_check}
	\begin{algorithmic}[1]
		\State \textbf{Inputs:} graph data $G$; ranking method $f$; node labels B;
        \State \textbf{Outputs:} overall SI vector \textbf{s}; positive/negative SI vector $\mathbf{s_{pos}}$/$\mathbf{s_{neg}}$; positive/negative SI vector in terms of labels $\mathbf{S_{pos}}$/$\mathbf{S_{neg}}$;
        \State $\mathbf{r_{original}} \leftarrow f(G)$
		\For {each node $v$ in $G$.nodes}
		\State remove $v$ and all connected edges \textbf{e} from $G$
		\State $\mathbf{r_{removed}} \leftarrow f(G)$
		\State calculate $\mathbf{r_{diff}}$ with $\mathbf{r_{original}}$ for each node in $ \mathbf{r_{removed}}$ 
		\For {each $r_{diff}[i]$ in $\mathbf{r_{diff}}$}
		\If {$r_{diff}[i] > 0$}
		    \State $s_{pos}^b[v] \leftarrow s_{pos}^b[v]  + abs(r_{diff}[i])$ for $b$ = label(i)
		\EndIf
		\If {$r_{diff}[i] < 0$}
		    \State $s_{neg}^b[v]  \leftarrow s_{neg}^b[v] + abs(r_{diff}[i])$ for $b$ = label(i)
		\EndIf
        \State $s[v] \leftarrow s[v] + abs(r_{diff}[i])$

		\EndFor
		 \State $s_{pos}[v] \leftarrow sum(s_{pos}^b[v])$; $s_{neg}[v] \leftarrow sum(s_{neg}^b[v])$ for $b$ in B
        \State add all $s_{pos}^b[v]$  to $S_{pos}[v]$ and all $s_{neg}^b[v]$  to $S_{neg}[v]$ for $b$ in B
		\State add $v$ and \textbf{e} back to $G$ 

		\EndFor
        \State \textbf{Return} $\mathbf{s}$, $\mathbf{s_{pos}}$, $\mathbf{s_{neg}}$, $\mathbf{S_{pos}}$, $\mathbf{S_{neg}}$
	\end{algorithmic} 

\end{algorithm} 

\vspace{2mm} \noindent  \textbf{Influence Overview:} The influence overview provides basic information on changes caused by removing a specific node (D2.1). These changes include 1) the number of influenced nodes which have ranking changes after the perturbation; 2, 3) the number of influenced nodes whose ranking increased/decreased after the perturbation; 4, 5) the max/min of increased/decreased ranking changes; 6, 7) the median of increased/decreased ranking changes; and 8) the degrees of the node. 
These 8 metrics provide the analyst with a statistical overview of the ranking influence of nodes due to perturbations. 
The radar chart is used to provide an overview of the sensitivity metrics with respect to the effects of a perturbation. (Figure~\ref{fig:teaser} (3)) The radar chart allows for the further addition of new metrics and can also preserve the overall information of the perturbation when the analyst switches between multiple perturbation diagnoses through the tabs on the top of the view.

\vspace{2mm} \noindent \textbf{Influence Distribution View (D2.2):} In addition to showing the snapshot of ranking changes, we also provide a ranking change distribution view and the top-$k$ proportional distribution view.

\begin{figure}[tbh]
    \centering	
    \includegraphics[width=1.00\columnwidth]{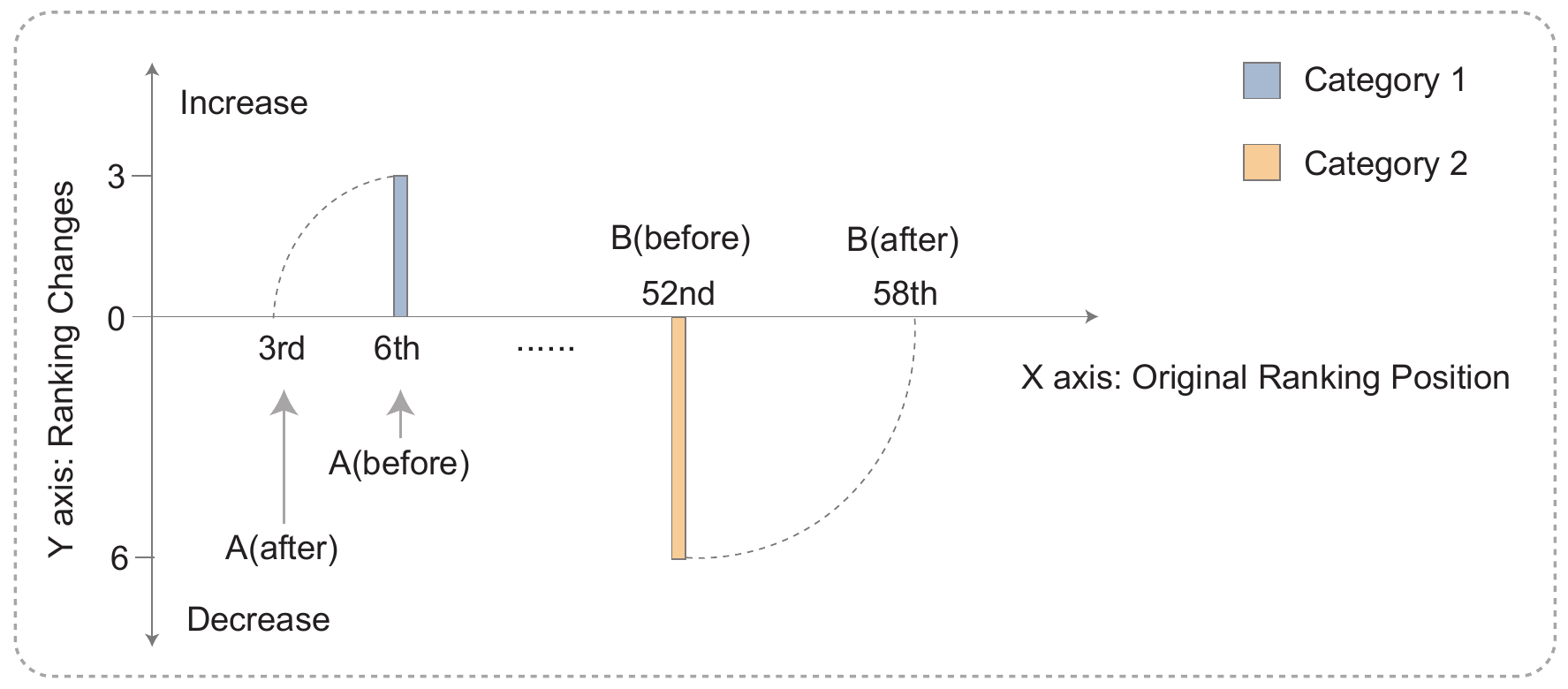}
    \caption{Visual encoding of the Ranking Change Distribution View. We use a bar chart to show the ranking change distribution. Each bar is a node. The position of the bar on the x axis denotes the original ranking position for the node. The height of the bar on the y axis denotes the ranking change for the node. Colors represent the node labels. We scale the axes of the bar chart such that a 90-degree clockwise rotation of the bar also allows the analyst to infer the future rank of the node.}
    \label{fig:ranking_change_distribution_view}
 	\vspace{0.3mm}
\end{figure}

\vspace{2mm} \noindent \emph{Ranking Change Distribution View}: A bar chart (Figure~\ref{fig:ranking_change_distribution_view}) is used to show the ranking change distribution.
Each bar is a node. The position of the bar on the x axis denotes the original ranking position for the node.
The height of the bar on the y axis denotes the ranking change for the node. 
Colors represent the node labels. We scale the axes of the bar chart such that a 90-degree clockwise rotation of the bar also allows the analyst to infer the future rank of the node. 

\vspace{2mm} \noindent \emph{Top-$k$ Proportional Distribution View}: Chartier et al.~\cite{10.1137/090772745} noted that when applying graph-based ranking algorithms for search engines, there could be an argument that ranking changes of webpages that are not part of the top-$k$ ranking are less important than those in the top-$k$. 
This argument can be extended to any general graph ranking problem, where the analyst can choose a $k$ for which elements below this ranking will not be considered.
For example, a hiring manager may not be interested in resumes ranked outside of the top-25, but it important to understand whether certain node attributes are underrepresented in the top-$k$.
In the Top-$k$ Proportional Distribution View, we use two donut charts to represent the proportions of nodes of different categories belonging to ranking 1 to ranking $k$ before and after the perturbation (Figure~\ref{fig:teaser} (5)), and $k$ is interactively specified.

\vspace{2mm} \noindent \textbf{Influence Detail View:} In the influence detail view, a data table is used to give the exact details about the node name, previous ranking, perturbed ranking, ranking difference, and labels. The analyst can sort all columns in the table, and, by hovering over a row, the location of the corresponding node will be highlighted in the local influence graph.

\vspace{2mm} \noindent \textbf{Influence Graph View:} While summarizing the changes in rank is important, our domain experts also required the ability to explore the impacts on the graph topology caused by perturbations. Note that for the graph-based ranking algorithm, such as PageRank and HITS, the underlying logic propagates the ranking value until convergence. In other words, the probability value for each node is propagated via a directed link connected between a node and its predecessors \footnote{A node that has a link that points to a given node in a directed path.}. In our case, perturbation is equivalent to removing the designated node and the links associated with it. As such, perturbation may cause two types of influence: \textit{direct influence} and \textit{indirect influence}. We consider direct influence to be the ranking changes of nodes due to the perturbation of their predecessors. Indirect influence is the ranking changes of nodes due to perturbations of any nodes that are not their predecessors. Understanding the direct and indirect influence is critical as analysts are interested in nodes that cause limited direct influence but have larger amounts of indirect influence.
These types of nodes are prime candidates for shadowing banning and other types of attacks as their removal can greatly influence the ranking results, but the removal will be relatively unnoticed by their direct connections, making such an attack difficult to quickly identify. 

To perform the direct/indirect influence analysis, we begin by building the influence graph which contains the topological relationships between the influenced nodes and the removed node. 
Here we introduce the concept of \textit{influence distance}, which we define as the geodesic distance \footnote{The length of the shortest directed path connecting the two nodes. For an unweighted graph, the length is the number of edges in the shortest path. The distance is \textit{infinite} if there is no path between two nodes.} between two nodes. We denote the influenced nodes, which are successors \footnote{Any node whose geodesic distance is equal to one from a path starting at $i$.} of the removed node, as \textit{hop-1} influenced nodes, which are also represented as the directly influenced nodes. Hop-1 influenced nodes have an influence distance of 1. 
Similarly, from the influenced nodes, we denote the hop-1 influenced nodes' successors as hop-2 influenced nodes, which have an influence distance 2, and so on and so forth. 
We define the influence distance to be the shortest distance between the removed node and the influenced node, which means that a hop-1 influenced node may also be a hop-3 influenced node's successor, but we only consider it as hop-1. 
Finally, for those nodes whose influence distance is infinite, we denote them as hop-inf influenced nodes. 
Algorithm~\ref{alg:influence_graph_construction} details our influence graph construction algorithm.

\setlength\textfloatsep{1.3mm}
\begin{algorithm}
	\caption{Influence Graph Construction}
    \label{alg:influence_graph_construction}
	\begin{algorithmic}[1]
		\State \textbf{Inputs:} graph data $G$; removed node $v_r$; influenced nodes $V'$ 
        \State \textbf{Outputs:} influence graph $G'$
        
        \State initialize $G'$
        \State initialize queue $\mathbf{q}$
        \State $G'$.addNode($v_r$, hop=0)
        \State $\mathbf{q}$.push($v_r$)
		\While {$\mathbf{q}$ is not empty}
		    \State $v \leftarrow \mathbf{q}$.pop()
		    
		    \For{$v_{neighbor}$ in $v$.neightbors()}
		        \If{$v_{neighbor}$ is influenced and not visited}
		        \State visited($v_{neighbor}$) $\leftarrow$ true
		        \State $G'$.addNode($v_{neighbor}$, hop=$v$.hop + 1)
		        \State $G'$.addEdge($v$, $v_{neighbor}$)
		        \State $\mathbf{q}$.push($v_{neighbor}$)
		        \EndIf
		    \EndFor
		\EndWhile
	    \If{any influenced node not in $G'$}
	        \For{$v_{remain}$ in remained influenced nodes}
	            \State $G'$.addNode($v_{remain}$, hop=inf)
		        \State $G'$.addEdge($v$, $v_{remain}$)
	        \EndFor
	    \EndIf
    \State \textbf{Return} $G'$
	\end{algorithmic} 
\end{algorithm} 

\begin{figure}[tbh]
    \centering	
    \includegraphics[width=1.00\columnwidth]{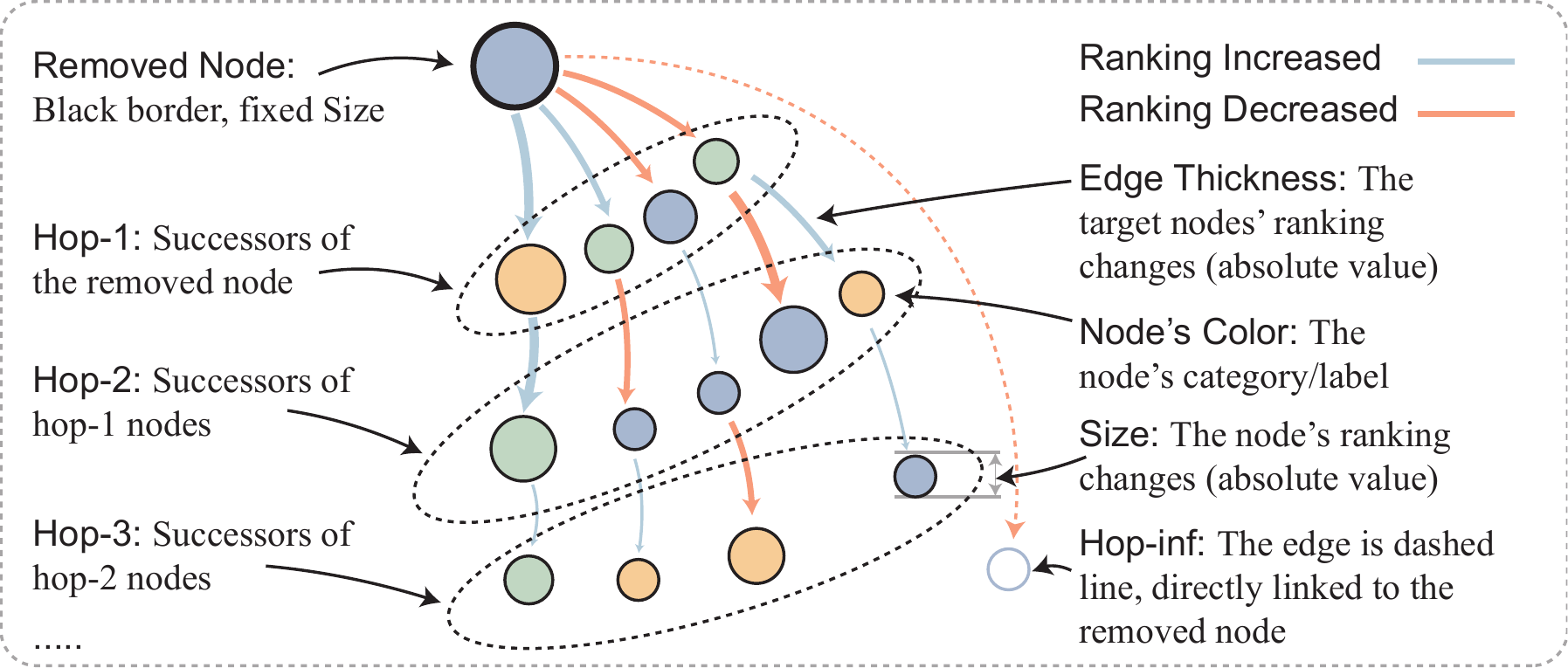}
    \caption{Visual encoding of the Influence Graph View. The perturbed node is placed at the left top of the graph and all successors which have ranking changes are grouped as hop-1 nodes. The successors of any hop-1 nodes which have ranking changes are grouped as hop-2 nodes, so on and so forth. The nodes' colors encode their categories (using colored filling and black strokes for regular nodes, and white filling and colored strokes for hop-inf nodes), and the nodes' sizes and thickness of their incoming edges encode the absolute value of ranking changes of nodes. The orange and light blue edge colors denote a ranking decrease/increase respectively. Nodes that do not have connections with any of the nodes in the influence graph are hop-inf nodes and are connected to the perturbed node with a dashed arrow line.}
    \label{fig:visual_encoding_of_local_influence_graph}
 	\vspace{0.5mm}
\end{figure}

We visualize the influence caused by removing/perturbing a node (D2.4) as a customized radial graph layout (Figure \ref{fig:visual_encoding_of_local_influence_graph}). In this customized layout, the removed node is set as the center of the force, and the strength of the charge force for each type of the node (hop-1 node, hop-2 node, etc.) is increased gradually based on the number $n$ of hop-n. In this way, all the nodes are clustered, and the influence graph forms a tree-like structure where the root of the tree starts from the top-left of the view and the branches spread towards the bottom-right of the view. Compared with a traditional force directed layout, this layout has two advantages:
1) All the influenced nodes are organized and positions of the nodes are relatively fixed in this layout, which enables the analyst to preserve their mental model. 
2) Nodes of the same type are clustered in this view, enabling the analyst to explore the composition for each cluster, i.e., each group of hop-n nodes. 
The color of the edges is encoded with light blue and orange, which shows whether the influenced node increased/decreased. 
The nodes' colors encode their categories (using colored filling and black stroke for regular nodes, and white filling and colored stroke for hop-inf nodes), and the nodes' sizes and thickness of their incoming edges encode the absolute value of the ranking changes of nodes.
We also provide interactive graph filtering so that analysts can filter by nodes whose rankings have increased/decreased, or nodes within certain influence distance ranges. The ranking change distribution view will also automatically update based on the ranges, Figure~\ref{fig:teaser} (6) (7). By filtering, the analysts can answer questions related to the perturbation, such as whether the node has a large direct influence on specific categories, or whether the node has a large indirect influence on nodes that are far away.

\subsection{Customized Constraints Filtering}
As we hinted at with the discussion of exploring ranking changes with respect to node attributes, in real-world applications, measures of sensitivity may need to be done with respect to domain specific constraints. Consider the Google search engine as an example, where each website is more concerned about reaching the first page of search results as well as climbing the ranking on that first page. It is reported that Google traffic is captured by 91.5\% of the first page search results~\cite{HowValua1online}, and there are also reports suggesting that top-3 results capture upwards of 75\% of the clicks~\cite{WeAnalyz3:online}. 
In such a setting, a domain owner would be interested in how sensitive their website is to ranking changes as a change in ranking from the first page of the search results to the second can have disastrous implications for web traffic. 

\begin{figure*}[tbh]
    \centering	
    \includegraphics[width=2.00\columnwidth]{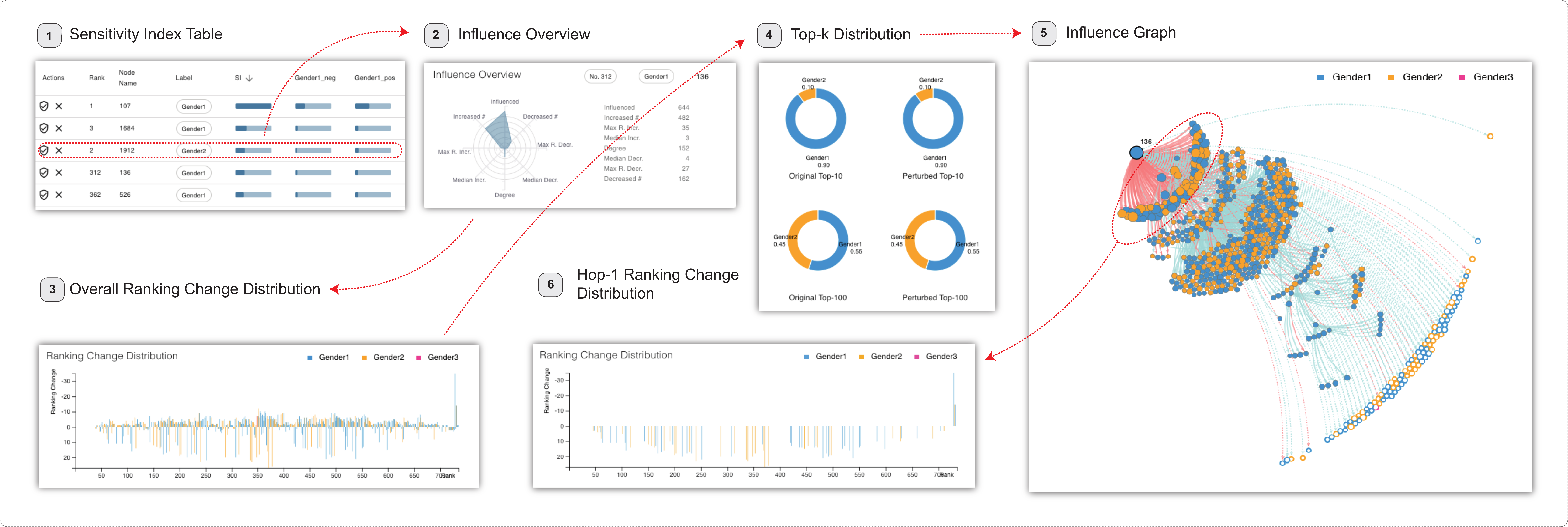}
    \caption{Facebook sensitivity analysis on PageRank. (1) The sensitivity index list shows that  \textit{user 136} has the 4th largest influence on the Sensitivity Index while its ranking score is 312 out of 734 nodes, which is considerably low for such a large influence. (2) The influence overview shows that the removal of  \textit{user 136} influences 644 of the 734 nodes, with 482 of them being positively influenced (ranking increased) while 162 are negatively influenced (ranking decreased). (3) The ranking change distribution view shows that negatively influenced nodes see a larger ranking decrease in comparison to positively influence nodes. (4) The top-k distribution view shows that as k increases from 10 to 100, the proportion of gender 2 increases, which means nodes of gender 1 are ranked relatively higher than nodes of gender 2.
    (5) The influence graph view further explains that most of those nodes who have large ranking changes are the neighbors of the removed  \textit{user 136}, and (6) the corresponding distribution view shows that those neighbors' rankings are evenly distributed. }
    \label{fig:case3}
	\vspace{-4mm}
\end{figure*}

Our framework enables customized constraints filtering functionality that allows analysts to add constraints to the sensitivity index list. 
Specifically, the analyst can define constraints that prevent selected nodes' rankings from increasing/decreasing by a certain degree. 
As Figure~\ref{fig:visual_analytics_framework} (C) shows, the analyst may have domain knowledge of the data and may wish to add constraints to the sensitivity index list to filter out any perturbations that would violate the constraints.
For example, if the analyst wants all the possible perturbations on the sensitivity index list that do not cause the top-3 nodes to experience ranking drops, the analyst can add a constraint: prevent top-3 nodes from ranking decreased by 0. The analyst can then sort the sensitivity index list and add the top-3 ranked nodes to the protected list by clicking the shield-like button for each of them and then configure a new rule ``protect selected nodes from ranking decreased by 0''. Finally, the analyst clicks the \textit{Update Constraints} button to add the new rule. The newly configured rule will then be displayed on the \textit{Rules} section (Figure~\ref{fig:teaser} (1)) and the sensitivity index list will be automatically updated such that any potential perturbations in it will not result in the top-3 ranked nodes having their rank decreased.
The analyst can also add more constraints as they may find certain nodes need to be protected from the perturbation during the diagnosis process. For example, if the analyst finds that a perturbation causes a significant ranking drop on the node, the analyst can use the lasso tool to select the node in the influence graph view (the node with significant ranking changes is encoded as a large circle in this view), and add it to the protected nodes and then apply a new rule. The analyst can then restart exploring the potential perturbations that do not violate the new rule. In this way, the analyst is able to explore possible perturbations under a variety of customized constraints.

%% file: 5_case_study.tex
\raggedbottom
\section{Case study and expert review}
In this section, we present three case studies to demonstrate how our framework supports sensitivity auditing for graph-based ranking. 
We showcase how data scientists analyze the sensitivity of the PageRank algorithm on Facebook social network data, how ranking developers check the robustness of their graph-based ranking algorithms, how social scientists analyze the exposure of blogs, and how the sensitivity of ranking algorithms can help identify potential manipulations.

\subsection{Facebook Ranking with PageRank}
In social network analysis, ranking members based on the graph structure is essential to tasks such as advertising~\cite{heidemann2010identifying}, social link prediction~\cite{gleich2015pagerank}, and recommendation~\cite{Gori:2007:IRB:1625275.1625720}. Perturbations in rankings may have a significant influence on the related business strategies. As such, it is important to audit the sensitivity of such rankings as this may help uncover malicious accounts, or reveal unintended biases in the algorithm. For example, analysts employ graph-based ranking methods to provide recommendations within a social network. These recommendations are based on the ranking results which predict who in the social network might be interested in the recommended products. Here, it may be important to understand if one sub-population is being over-targeted with particular advertisements. In this case study, we analyze ranking sensitivity in the Facebook social network dataset~\cite{leskovec2012learning}. For demonstration purposes, the social network is down-sampled into a graph with 734 nodes and 74254 edges. We use the gender of each network member (where each member is a node in the graph) as the class label and explore if perturbations in the network reveal ranking bias with respect to gender. 

\vspace{2mm} \noindent \textbf{Identifying the Instance-level Sensitivity (T1):} The sensitivity index table is displayed after loading the graph data and choosing the ranking method, Figure~\ref{fig:case3}. We sort the column \textit{SI (Sensitivity Index)} in order to explore which node (network member) is likely to have the largest influence on the ranking result. After sorting by SI, it can be observed that the nodes with the highest SI are also the highest ranked nodes. This phenomenon matches the explanation in Kang et al.~\cite{kang2018aurora} that the nodes with high influence are also often highly ranked. 
Interestingly, though, we see that the 4th node,  \textit{user 136}, has a high sensitivity index. However, the network member is ranked 312th out of 734 nodes. This particular case is of keen interest to our analyst.

\vspace{2mm} \noindent \textbf{Diagnosing the Perturbation Effects (T2):} By clicking `diagnose the perturbation effect' on  \textit{user 136}, the details of the influence is depicted in the influence overview. In Figure~\ref{fig:case3} (2), the influence overview shows that the perturbation caused by removing  \textit{user 136} has influenced 644 out of 734 nodes, where 482 nodes' rankings increased and 162 nodes' rankings decreased. By further exploring the ranking change distribution view, we find that although positively influenced nodes are 3 times more common than the negatively influenced nodes, the negatively influenced nodes are subject to larger ranking fluctuations than the positively influenced nodes. While we expect that the nodes that were previously ranked lower than  \textit{user 136} would see a rise in ranking to fill the gap created by  \textit{user 136}, it is surprising to also find large negative changes occurring. 
In the ranking distribution view and the top-$k$ distribution view, Figure \ref{fig:case3} (3), we observe that the perturbation does not result in drastic changes in the ranking distributions.
Furthermore, if we split the top-$k$ rankings by gender, the perturbation is not observed to impact one gender class's rankings more than another.

In addition, our analyst is interested in the relationships between the ranking changes and the topological structure. 
Specifically, the analyst wonders how the removed  \textit{user 136} is connected to the influenced nodes given the stronger impacts for decreased ranking. 
In the local influence graph view, the nodes who have significant ranking changes in the perturbation are directly connected to  \textit{user 136} since all the large circles are hop-1 node, Figure~\ref{fig:case3} (4). 
After selecting the range of influence distances, we can see that the decreases of rankings for most of the nodes are caused by being the immediate successor of \textit{user 136}. 
Furthermore, most of the nodes outside the hop-1 circle have their ranks slightly increased. 
This may be due to the fact that those nodes are not heavily influenced by the perturbation.
\begin{figure*}[tbh]
    \centering	
    \includegraphics[width=2.00\columnwidth]{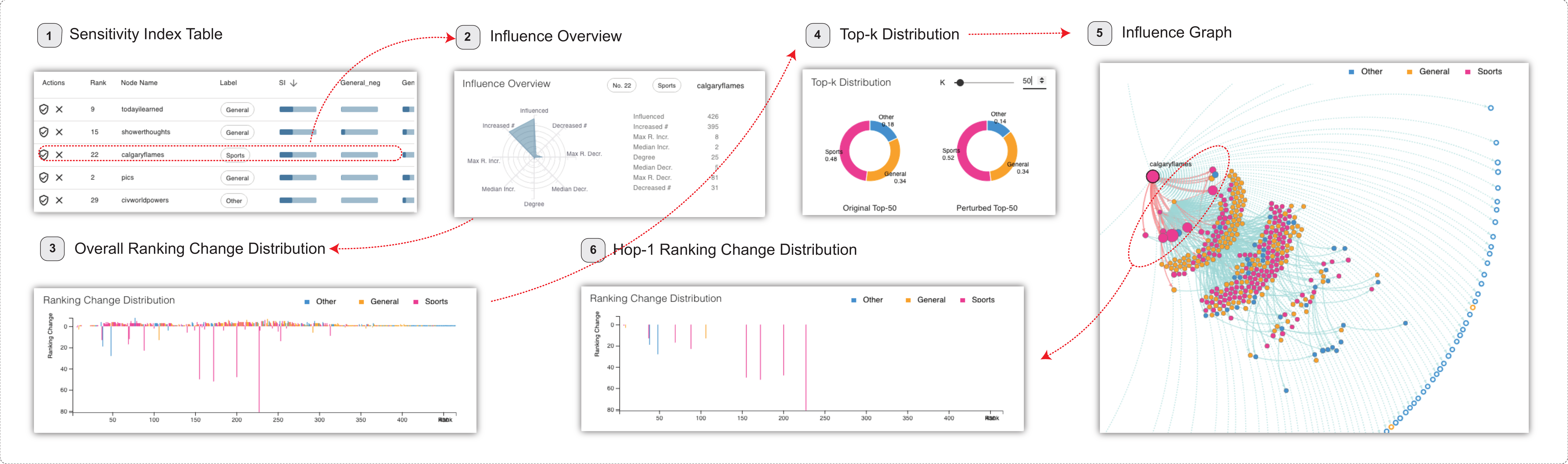}
    \caption{Subreddit sensitivity analysis on PageRank. (1) The sensitivity index list shows that the \texttt{r/CalgaryFlames} node belongs to the sports topics and has a large sensitivity index. (2) The influence view shows that removal influenced 426 out of 464 nodes, and 395 of them are positively influenced while only 31 are negatively influenced. (3) The ranking change distribution view shows that a few nodes experience a large ranking decrease due to the removal of \texttt{r/CalgaryFlames}, the majority of which are sports topics. (4) The top-k distribution view shows that the subreddits for sports occupy more of the top-50 ranks than other subreddit topics. (5) (6) The removed node has relatively few neighbors that are negatively influenced, and there are more direct influences in the `General' nodes than `Other' nodes.}
    \label{fig:case2}
	\vspace{-4mm}
\end{figure*}

\subsection{Subreddit Ranking with PageRank}
In the second case study, we audit the sensitivity of PageRank results on the subreddit community interaction graph dataset~\cite{ilprints422}. 
Kumar et al.~\cite{10.1145/3178876.3186141} studied the community interactions and conflicts between communities in Reddit to show that a community can be mobilized by negative sentiment comments from another community. 
Such conflicts between communities can potentially reduce the activities among community members and may lead to people leaving the platform. 
In such cases, it is also possible that other communities will be influenced due to chain effects in the network, which may cause member churn and increased complaints about the platform. Thus, the Reddit community managers and data analysts may be interested in inspecting the activities of communities to make sure the content environment is benign and also be aware of any perturbations (deletion of subreddit posts/activity reduction) that might influence ranking results during any possible recommendation processes.  

In this case, we sub-sampled the subreddit community interaction dataset~\cite{10.1145/3178876.3186141} down to 464 nodes and 6676 edges. 
Each node represents a subreddit, and a topic label is assigned to the nodes to identify their main categories, such as \texttt{r/basketball}, \texttt{r/soccer} and \texttt{r/nfl} under the topic ``Sport''. 
An edge between two nodes indicates there is a comment in one subreddit which referred to the other subreddit in the content of the comment. 
We select three topics to explore: Sports, General, and Others.
In this case, comments with a high page rank would receive more views, and if negative comments generate more clicks, this could be of concern.
Community administrators could utilize our framework to explore the interactions between communities and identify potential issues of flaming, karma farming, etc.

\vspace{2mm} \noindent \textbf{Identifying the Instance-level Sensitivity (T1):}
After the ranking results are loaded into the system, we want to know which subreddit in the ``Sports'' topic has the largest influence on the reputation of other subreddits. By sorting the sensitivity index column, the ranking result is listed in the sensitivity index table, and the subreddits are ordered by their sensitivity values in descending order. We believe that concrete entities such as sports players and teams can have more controversial topics, but are less noticeable by community members who are not interested in them. Thus, we skip the general subreddits such as \texttt{r/hockey, r/baseball, and r/basketball} and focus on the \texttt{r/CalgaryFlames}, which has a relatively large influence on the reputation rankings of the subreddits.

\vspace{2mm} \noindent \textbf{Diagnosing the Perturbation Effects (T2):} The impact of removing the \texttt{r/CalgaryFlames} node are shown on the right side of Figure~\ref{fig:case2}. In the influence overview, the removal has influenced 426 out of 464 nodes. Among the 426 nodes, 395 of them have their ranks increased while 31 decreased. Specifically, the node \texttt{r/thebeach} has increased by 8 positions, and the rank of node \texttt{r/coloradoavalanche} has dropped by 81 positions. This indicates that the perturbation has triggered massive declines even though the average increase is relatively low. In the ranking change distribution view, we observe that the ranking declines occur primarily in the Sports subreddits whose original ranks are relatively low (around 255 out of 464). Compared with the decrease, the overall distribution of the increased nodes covers a broader range on the original rankings; however, a much smaller climbing effect in the ranking positions is observed. This may be due to the fact that the original rank of the \texttt{r/CalgaryFlames} is very high (22nd), which could possibly bump many lower-ranked subreddits into higher positions. In the distribution view, all three topics (sports, general and others) receive slight increases in the median values. However, the overall distribution remains the same. We further query the top-50 since there are considerable ranking changes depicted in the ranking change distribution view. This suggests that after removing the \texttt{r/CalgaryFlames}, the proportion of subreddits in the category of Sports among the top-50 has increased from 48\% to 52\%, while the proportion of ``Other'' subreddits has dropped from 18\% to 14\%. In the influence graph view, we find that the removal results in large drops the to ranks of the hop-1 nodes, which matches the patterns shown in the ranking change distribution view. That is to say, the removal of \texttt{r/CalgaryFlames} only influences its neighbors with significant ranking changes. By filtering on the influence distance values, we find that the perturbation significantly influences the ``Sports'' subreddits.

\subsection{Political Blogs Ranking with HITS}
Graph-based ranking methods are widely used to rank webpages. Here, we consider a scenario where removing certain pages from a website (either intentionally by the website owner, or maliciously through shadow bans by an external party) can significantly change the rankings of other pages. Consider a political web forum where members post views and opinions on certain topics and issues. The search result rankings are determined by a graph ranking algorithm, and higher-ranked opinions are more likely to be read and shared. Here, one could imagine a nation-state actor that would want to promote biased content. The nation-state actor can create webpages to add various links in the graph structure, or even identify websites to shadow ban, which could manipulate the ranking results so that certain political opinions are more exposed to the public. 
By using the articles in the forum as the nodes and the hyperlinks between different articles as edges, the graph ranking methods can recommend popular articles in the forum based on the graph structure. However, there could be some nodes that are vulnerable and have high sensitivity indices with respect to the graph ranking method. They become the target for the attackers who wish to manipulate the article rankings and promote their own content. As such, blog managers and social scientists may wish to collaborate with ranking algorithm developers to make sure such ranking results are fair and stable with respect to potential perturbations (deletion of blogs).

In this case study, we explore the sensitivity of the HITS algorithm on the political blog dataset~\cite{10.1145/324133.324140}. The dataset includes a topic citation graph between liberal and conservative blogs prior to the 2004 U.S. Presidential Election. We subsampled 397 nodes (i.e, blogs) and 12,365 edges (i.e., hyperlinks between blogs), removing all nodes with degree less than 30. Our goal was to explore the structural changes in the network that result in drastic ranking changes in the HITS algorithm.

\vspace{2mm} \noindent \textbf{Identifying the Instance-level Sensitivity (T1, T3):} Before the analysis, we made three assumptions about ranking manipulation: 1) the top-$k$ items in the ranking are much more important than other nodes since readers typically only view the top results provided by the search engine. 2) It is riskier to manipulate a node with a higher rank since the readers may notice the changes. 3) To avoid having manipulations discovered, the attacker would assume a posture of minimum risk. As such, our goal is to discover how we can manipulate the ranking results by removing a node, while working under these constraints.

Based on our constraints, we apply selection rules to filter the sensitivity index table. We first click the ranking column to sort the ranking order. Then, we add the top-5 nodes to the protected nodes with constraints of \textit{protect selected nodes from their ranking decrease by 0\%}, which excludes all perturbations that would cause the rankings of these selected nodes to decrease. The constrained sensitivity index table is shown on the left, which contains 1) the ranking positions, 2) node names, 3) overall sensitivities, and 4) sensitivity details including positive/negative influence to liberal/conservative blogs. After sorting the rows by the sensitivity index column in decreasing order, a liberal blog, \texttt{liberaloasis.com} appears in the second row of the table. By observing the other columns, we find that \texttt{liberaloasis.com} is not in the top-10 rank; however, its removal can increase the rankings of the conservative blogs while decreasing the rankings of the liberal blogs. 

\vspace{2mm} \noindent \textbf{Diagnosing the Perturbation Effects (T2):} Next, we explore why this blog is so influential. After selecting ``Explore the Perturbation in Detail'' by clicking the cross button in the first column, all the details are listed on the right side of the interface (Figure~\ref{fig:teaser}). In the influence view, the radar chart indicates that there are 368 out of 397 nodes influenced by the removal of \texttt{liberaloasis.com}. 232 of the 368 nodes have their ranks increased while 136 decreased. The largest increase in the ranking positions is 16, and the largest decrease is 30. From the ranking change distribution (Figure~\ref{fig:teaser} (6)), we find that most of the ranking changes happen in the range between 50 and 150. 
Since only mid-tier ranks are impacted, the effects of removing this node are subtle, meaning this change is not easily observable. 
However, the impact is significant. 
In the distribution changes view, we can observe that the median of the liberal blog ranking distribution decreases, while the conservative blog ranking distribution increases. 
By further exploring the proportions of both liberal and conservative blogs in the top-100 results, Figure~\ref{fig:teaser} (5), we can see that the proportion of liberal blogs in the top-100 results has decreased by 5\% (from 82\% to 77\%), thus subtlety shifting the site's content.

In the ranking change distribution view, we also find that there are three liberal blogs with considerable ranking decreases after the perturbation. We further check the detailed view for information on the influenced nodes. We sort the original column to locate the exact ranking position and notice that the first three liberal blogs, \texttt{sununes}, \texttt{boloboffin}, and \texttt{elemming2}, have a large ranking decrease after the perturbation, which corresponds to the three liberal blogs in the ranking change distribution view mentioned above. We want to further explore the relationship between the ranking changes with the topological structure. The influence graph view (Figure~\ref{fig:teaser} (8)) shows that the removed node influences a majority of liberal nodes, and as the influence distance increases, conservative nodes are indirectly influenced. 

\subsection{Expert Review}
Along with our case studies, we have conducted a group interview with our collaborator (E0) and three additional domain experts in graph mining (E1, E2, and E3) to provide feedback on our framework. We began the interview by introducing our visual analytics framework, and the functionalities supported by each module. Then, we presented a demo of the analytical flow across the three previously described case studies. After this, the experts are allowed to freely explore the perturbation results of the three datasets (Facebook, Reddit, and Polblogs) over two ranking methods (Pagerank and HITS) in our system. The interview lasted approximately 90 minutes, and we collected free-form responses to the following questions: 
\begin{enumerate}[topsep=0.3em,partopsep=0em,parsep=0.2em,itemsep=0em,leftmargin=*,labelsep=0.4em]
    \item Does the system meet the design requirements and address the analytical tasks proposed in our work? 
    \item Does our analytical pipeline match your daily workflow?
    \item How is the information delivered through our system?
    \item How would you perform the same tasks in conventional graph mining methods? 
\end{enumerate}

\noindent \textbf{Framework:} We received positive feedback from the experts in terms of our proposed framework. The experts found the framework to be practical with respect to the proposed problems. E1 appreciated that the framework is capable of handling the sensitivity issues that are related to nodes' attributes, and noted that such a framework can support not only graph-ranking developers but also experts in other fields evaluating whether the ranking algorithm is suitable for real-world ranking tasks. E2 appreciated our constraint filtering functionality since such an iterative analysis is one of his preferred approaches.

\noindent \textbf{Visualization:} The overall visualization techniques were also well received. E1 mentioned: \textit{``With the knowledge of understanding what the system is capable of doing, I found most of the visualizations are straightforward and easy to understand. The newly designed influence graph is also intuitive once I learned what each encoding means.''} E2 also appreciated that the interactive effect is helpful for understanding the ranking change effects after the perturbation. E3 suggested that we could add question-mark icons that link to descriptions for each view.

\noindent \textbf{Limitations:} The experts also identified several limitations of our current framework. E1 and E2 noticed that there is only one perturbation method supported in our system, which is the node removal. However, they all understood that the perturbation spaces (edge removal, node/edge addition) are far larger than the node removal space. E3 also mentioned that the visualizations may become crowded when the size of the influence graph is large.

%% file: 6_conclusion.tex
\section{Conclusions and future work}
In this work, we propose a visual analytics framework for auditing the sensitivity of graph-based ranking methods. By analyzing the influence of the ranking method due to a node's removal, analysts can diagnose the perturbation effects in terms of ranking changes. The system is targeted for graph ranking developers and graph-related domain experts.
The framework is implemented using D3 for the visual components and Python 3 (the NetworkX library~\cite{SciPyProceedings_11}) for data processing. All source code is provided in Github~\footnote{https://github.com/VADERASU/auditing-sensitivity-graph-ranking}.

\vspace{2mm} \noindent \textit{Scalability:} The computation of the initial sensitivity check is the major bottleneck of our framework. This process requires re-computation of ranking results, calculating the ranking changes of all nodes, and pre-computing the statistical data for each perturbation. 
Even though we have pre-computed the required data for the visualization to mitigate the real-time computation burden, the overall preparation time of the initial sensitivity check is $O(n^2m)$, where $n$ is the number of nodes and $m$ is the number of edges. In our case studies, it takes approximately 2 minutes 30 seconds to pre-compute the Reddit dataset with 464 nodes and 6676 edges, 3 minutes for the Polblogs dataset with 397 nodes and 12365 edges and 25 minutes 30 seconds for the Facebook dataset with 734 nodes and 74254 edges. When a large graph is loaded, the computation time for processing the data can be prohibitively long on commodity hardware. This is the reason that we decided to subsample our two datasets instead of exploring the whole dataset. The corresponding effect is that the ranking result and sensitivity result may be different from the real dataset. Future work will explore parallel solutions and novel methods for further sensitivity calculations. 

\vspace{2mm} \noindent \textit{Visual Design:} 
The visual design for the sensitivity index list is a bar representation for each sensitivity column. The advantage is that the analyst can sort every column, which allows them to quickly identify sensitivity aspects of interest. However, when a node has multiple labels, the sensitivity check for those aspects must also be displayed as columns, which consumes more space and reduces the ability of the analyst to quickly browse the results. Alternatives include re-designing the sensitivity check into glyphs so that the analyst can compare patterns and we will save display real estate. Another design issue occurs in the influence graph view, where the edge length is defined as the influence distance. Here we only consider a fixed number of influence distances, 9. This means that for large graphs, the display space will be limited.% and information could be lost. 

\vspace{2mm} \noindent \textbf{Future work: }
In this work, we only audit sensitivity based on one perturbation strategy, node removal. In reality, there are also other types of perturbations of the graph, including adding/removing edges and adding nodes, where manipulating edges can be considered as establishing a connection/follow/like/cite other nodes, and adding nodes is a more subtle way of perturbation since it is often easier to create an account/website/user, etc. However, adding graph elements is computationally expensive as additions make the perturbation space infinite. Possible solutions would be  limiting the perturbation space to a constrained budget. However, it is important to note that the addition of graph elements could be supported within the proposed framework.
Our future work will target those types of perturbations.